\documentclass[10pt]{iopart}
\usepackage{graphicx,amssymb,amsfonts,units,bm,amsthm,hyperref} 

\begin{document}

\title{A Non-Commutative Formula for the Isotropic Magneto-Electric Response}

\author{Bryan Leung}

\address{Department of Physics \& Astronomy, Rutgers University, Piscataway, NJ 08854, USA \\
\href{mailto:bryleung@physics.rutgers.edu}{bryleung@physics.rutgers.edu}}

\author{Emil Prodan}

\address{Department of Physics, Yeshiva University, New York, NY 10016, USA \\ 
\href{mailto:prodan@yu.edu}{prodan@yu.edu}}

\begin{abstract}
A non-commutative formula for the isotropic magneto-electric response of disordered insulators under magnetic fields is derived using the methods of non-commutative geometry. Our result follows from an explicit evaluation of the Ito derivative with respect to the magnetic field of the non-commutative formula for the electric polarization reported in Ref.~\cite{Schulz-BaldesArxiv2012gh}. The quantization, topological invariance and connection to a second Chern number of the magneto-electric response are discussed in the context of 3-dimensional, disordered, time-reversal or inversion symmetric topological insulators.  
\end{abstract}


\date{\today}

\maketitle

\section{Introduction}

The magneto-electric effect in insulating materials consists in the appearance of a finite, measurable electric bulk polarization ${\bm P}$ when a sample is subjected to an external magnetic field ${\bm B}$, and in the appearance of a finite, measurable bulk magnetization ${\bm M}$ when the sample is subjected to an external electric field ${\bm E}$. The magneto-electric effect is quantified by the magneto-electric response tensor:
\begin{equation}
\alpha_{ij} = \frac{\partial{P_i}}{\partial B_j} = \frac{\partial{M_j}}{\partial E_i},
\end{equation}
where the derivatives are not necessarily taken at zero ${\bm B}$ or ${\bm E}$ fields. The effect has been observed in a variety of materials and its technological applications can be tremendous \cite{FiebigJPD2005ey,SpaldinScience2005tr}.

Originally, the effect was sought in materials with broken time-reversal and inversion symmetries, but recently it was shown that certain classes of time-reversal or inversion invariant systems can display large magneto-electric responses \cite{QiPRB2008ng,TurnerPRB2012cu,Hughes2010gh} due to topological characteristics of their electronic structures. Such materials, which are now routinely synthesized and characterized in laboratories \cite{ZHassanRevModPhys2010du,QiRMP2011tu}, are called 3-dimensional topological insulators. It emerged in the recent years that the 3-dimensional topological insulators can be classified by their magneto-electric response. For example, for perfectly periodic, time-reversal invariant insulators, the magneto-electric response induces a $\mathbb{Z}_2$ topological classification \cite{QiPRB2008ng} and the topological part of the magneto-electric tensor, which can serve as the topological invariant for this classification, was shown \cite{Wang2010xs} to equal the previously introduced \cite{Fu:2007ti,Moore:2007ew,Roy:2009am} $\mathbb{Z}_2$ invariants. This was an important development in the field of topological insulators because it provided a measurable bulk effect which sets apart the trivial and topological 3-dimensional time-reversal invariant insulators \cite{LiNature2010cm}.

The magneto-electric response has ionic, spin and orbital contributions. Here we discuss only the orbital component, which, at its turn, can be decomposed into isotropic and traceless parts:
\begin{equation}
\alpha_{ij}=\delta_{ij}\frac{1}{3}\sum_{k=1}^3 \alpha_{kk} + \tilde{\alpha}_{ij}.
\end{equation}
Both components are experimentally measurable quantities but we will focus exclusively on the isotropic component:
\begin{equation}
\alpha = \frac{1}{3}\sum_{j=1}^3 \alpha_{jj},
\end{equation}
for it contains the  topological response. For perfectly periodic, time-reversal or inversion invariant crystals at zero external fields, $\alpha$ was derived and explicitly computed in Refs.~\cite{QiPRB2008ng, EssinPRL2009bv,Hughes2010gh} for several lattice models of time-reversal or inversion symmetric insulators, showing that indeed $\alpha$ can take non-trivial quantized values. For perfectly periodic systems without time-reversal and inversion symmetries, Refs.~\cite{EssinPRB2010ls,MalashevichNJP2010bv} showed that $\alpha$ has additional contributions besides the topological component. 

Our goal here is to extend the above mentioned works to the case when disorder and magnetic fields are present, maintaining the assumption of a gap in the energy spectrum at the Fermi level. The main challenges arising in this more general setup are the breaking of the translational symmetry and the tedious linear response calculus with respect to the magnetic field. Both challenges are overcome with the use of non-commutative geometric methods \cite{Connes:1994wk}. The starting point for our work is Ref.~\cite{Schulz-BaldesArxiv2012gh} where non-commutative formulas for ${\bm P}$ and ${\bm M}$ have been reported. Given these results, we could have proceeded along two routs, one relying on the variation of ${\bm P}$ with ${\bm B}$ and the other one relying on the variation of ${\bm M}$ with ${\bm E}$. We chose the first route because computing the variation of ${\bm P}$ with ${\bm B}$ fits better into the non-commutative geometry of the Brillouin torus \cite{BellissardLNP1986jf},  which provides a set of explicit rules of calculus, including rules for how to compute the derivatives with respect to the magnetic field (see Sections~\ref{NCB} and \ref{Ito}). These rules of calculus alone enabled us to complete the task at hand.

For disordered time-reversal or inversion symmetric insulators (${\bm B}=0$), under the spectral gap assumption, we are able to formulate the following statement (similar to the existing one for perfectly periodic crystals \cite{QiPRB2008ng,TurnerPRB2012cu,Hughes2010gh}): The total change $\Delta \alpha[\gamma]$ of $\alpha$, induced by an adiabatic deformation along a path $\gamma$ in the model's parameter space, is equal to half of the second Chern number over the manifold $(\gamma-\Sigma \gamma)\times \mathrm{Brillouin \ torus}$ (when the units are chosen such that $e=h=1$). Here, $\Sigma$ is either the time-reversal operation or the inversion operation and the path $\gamma$ starts and ends with a time-reversal or inversion symmetric system. Since $\Delta \alpha[\gamma]$ can change only by an integer number if the path $\gamma$ is re-routed, it follows that the integer or half-integer character of $\Delta \alpha[\gamma]$ is independent of the deformation path and this extends the well-known $\mathbb{Z}_2$ topological classification to the disordered case. We would like to mention that the methods of non-commutative geometry enabled us to derive this result directly in the thermodynamic limit and to avoid the use of arguments based on twisted boundary conditions on large but finite volumes. The latter arguments are hard to control in the thermodynamic limit and they are problematic when strong disorder is considered.

The explicit formula for $\alpha$ (Eq.~\ref{MainFormula}) and its topological properties discussed in the present work is a small step in a program that aims at elucidating what happens when strong disorder is present. By strong disorder we mean situations when the Fermi level is embedded in a dense localized spectrum rather than being positioned in a spectral gap. The crucial question to answer is if the isotropic magneto-electric response remains quantized and invariant under the assumption of a mobility gap (as opposed to a spectral gap) at the Fermi level. A rigorous answer to this question will finally resolve the problem of existence (or lack) of robust extended states in 3-dimensional topological insulators.

\section{Disordered lattice-models in the presence of magnetic fields}\label{Model}

The analysis reported in this work applies to generic disordered lattice-models in $3$-dimensions. The Hilbert space for such a model is spanned by vectors of the form $|{\bm n},\alpha \rangle $, where ${\bm n}\in \mathbb{Z}^3$ is a node of the lattice and $\alpha=1,\ldots,D$ labels the orbitals associated with that node. The scalar product $\langle \cdot,\cdot \rangle$ is by definition such that $ \langle {\bm n},\alpha | {\bm m},\beta \rangle = \delta_{{\bm n}{\bm m}}\delta_{\alpha \beta}$. The generic lattice-Hamiltonians, in the absence of a magnetic field and disorder, take the form:  
\begin{equation}\label{LatticeHam0}
H_0=\sum_{{\bm n},\alpha}\sum_{{\bm m},\beta} t_{{\bm n}-{\bm m}}^{\alpha \beta} |{\bm n},\alpha \rangle  \langle {\bm m},\beta |,
\end{equation}
Accurate, material-specific lattice-models can be generated from {\it ab-initio} computations using Wannier function representations or simply by fitting the measured band structure \cite{LiuPRB2010xf}. In almost all cases, the lattice-Hamiltonians involve only a finite number of hoppings between one unit cell and its neighbors, which is something we will assume in the followings.

In the presence of disorder and a uniform magnetic field ${\bm B}$, which will be parametrized as a $3\times 3$ antisymmetric  tensor $\hat{{\bm B}}$:
\begin{equation}
\hat{{\bm B}}=\left (
\begin{array}{ccc}
0 & B_3 & -B_2 \\
-B_3 & 0 & B_1 \\
B_2 & -B_1 & 0
\end{array}
\right ),
\end{equation}
the lattice-Hamiltonian becomes:
\begin{equation}\label{LatticeHamO}
H_\omega=\sum_{{\bm n},\alpha}\sum_{{\bm m},\beta} e^{\mathrm{i}\pi({\bm n}, \hat{{\bm B}} {\bm m})}t_{{\bm n}-{\bm m}}^{\alpha \beta}(\omega) |{\bm n},\alpha \rangle  \langle {\bm m},\beta |,
\end{equation}
where the phase factor $e^{\mathrm{i}\pi({\bm n}, \hat{{\bm B}} {\bm n})}$ encodes the effect of the magnetic field through the Peierls substitution \cite{PanatiCMP2003rh}. Above, $(,)$ denotes the 3-dimensional Euclidean scalar product and $\mathrm{i}=\sqrt{-1}$. The hopping amplitudes now have a random component, with a typical form:
\begin{equation}
t_{{\bm n}-{\bm m}}^{\alpha \beta}(\omega)=t_{{\bm n}-{\bm m}}^{\alpha \beta}+\lambda \omega_{{\bm m},{\bm n}},
\end{equation}
where $\omega_{{\bm m},{\bm n}}$ are independent random variables uniformly distributed in the interval $[-\frac{1}{2},\frac{1}{2}]$. The collection of all random variables $\omega=\{ \omega_{{\bm m},{\bm n}}\}_{{\bm m},{\bm n}}$ can be viewed as a point in an infinite dimensional configuration space $\Omega$, which can be equipped with a probability measure:
\begin{equation}
dP(\omega)=\prod_{{\bm m},{\bm n}}d\omega_{{\bm m},{\bm n}}.
\end{equation}
Furthermore, the discrete $\mathbb{Z}^3$ additive group has a natural action on $\Omega$ given by:
\begin{equation}
(\mathfrak{t}_{\bm a} \omega)_{{\bm m},{\bm n}}=\omega_{{\bm m}+{\bm a},{\bm n}+{\bm a}},
\end{equation}
which acts ergodically on $\Omega$ and leaves $dP(\omega)$ invariant. If $U_{\bm a}$ denotes the magnetic translation by ${\bm a}$:
\begin{equation}
U_{\bm a}|{\bm n},\alpha \rangle = e^{ \mathrm{i} \pi ({\bm a}, \hat{{\bm B}} {\bm n})}|{\bm n}-{\bm a},\alpha \rangle,
\end{equation}
 then $\{H_\omega\}_{\omega \in \Omega}$ forms a covariant family of operators in the sense that:
\begin{equation}
U_{\bm a}H_\omega U_{\bm a}^{-1}=H_{\mathfrak{t}_{\bm a}\omega}.
\end{equation}

In the above settings, the quadruple $(\Omega,dP(\omega),\{\mathfrak{t}_{\bm a}\}_{{\bm a}\in \mathbb{Z}^3},\{H_\omega\}_{\omega \in \Omega})$ determines a homogenous system \cite{BellissardLNP1986jf}, for which the intensive thermodynamic variables are, with probability one, independent of the disorder configurations. The results reported in this paper can be extended to more general models of homogeneous systems, but one should keep in mind the model of Eq.~\ref{LatticeHamO} as a prototypical example.

We end this section with a comment about discrete versus continuum models. Definitely, working with lattice models reduces tremendously the technical complexity of the problem. One crucial difference is that the algebra of observables does have a unit in the discrete case, and it doesn't in the continuum case. Also, when working with discrete models, we are dealing only with bounded observable but this will not be the case for continuum models. Hence, extending the theory to the continuum models will require more work at the technical level (note also  that the results of Ref.~\cite{Schulz-BaldesArxiv2012gh}, which is starting point for our work, were established for discrete models). However, even if we do extend the theory to continuum models, an explicit computer calculation for a real material will not be possible, simply because of the sheer complexity of the problem. This is the main reason the community working on topological materials focuses almost entirely on lattice models. We should mention that, at least for the periodic case, the Wannier function technique generates lattice models which exactly reproduce the bundle of the occupied states of the continuum models. 

\section{The non-commutative Brillouin torus (Bellissard \cite{BellissardLNP1986jf})}\label{NCB}

\subsection{The $C^\ast$-algebra of the covariant observables}

If for perfectly periodic systems one is primarily concerned with the translationally invariant operators, for homogenous systems one is primarily concerned with the covariant families of operators. All relevant response functions and thermodynamic functions for a homogeneous model can be computed within the algebra generated by the covariant families of operators. This algebra can be given a $C^\ast$-algebra structure in the following way.

For $\{F_\omega\}_{\omega \in \Omega}$ a covariant family of operators, all the information is encoded in the matrix elements:
\begin{equation}\label{MatF}
f_{\alpha,\beta}(\omega,{\bm n})=\langle {\bm 0},\alpha | F_\omega |{\bm n},\beta \rangle,
\end{equation}
because all the other matrix elements can be obtained from $f_{\alpha,\beta}(\omega,{\bm n})$ using translations. Hence, instead of working with covariant families of operators, one can work with functions:
\begin{equation}
f:\Omega \times \mathbb{Z}^3 \rightarrow {\cal M}_{D \times D},
\end{equation}
where ${\cal M}_{D \times D}$ is the space of $D \times D$ complex matrices, and with the algebraic operations that follows from the operator algebra when written for the matrix elements of Eq.~\ref{MatF}. These operations are:
\begin{equation}\label{AlgRules}
\begin{array}{l}
(f+g)(\omega,{\bm n})=f(\omega,{\bm n})+g(\omega,{\bm n}), \medskip \\
(f*g)(\omega,{\bm n})=\sum_{{\bm m} \in {\mathbb Z}^d} f(\omega, {\bm m})g(\mathfrak{t}_{{\bm m}}^{-1}\omega,{\bm n}-{\bm m})e^{\mathrm{i} \pi( {\bm n}, \hat{{\bm B}} {\bm m})}.
\end{array}
\end{equation} 
For the beginning, one should consider only continuous functions $f$ with compact support. The algebra generated by such elements will be called ${\cal A}_0$. One can easily go back from the algebra ${\cal A}_0$ to the algebra of bounded operators using the operator representation $\pi_\omega$ given by the formula:
\begin{equation}\label{OpRep}
(\pi_\omega f)|{\bm n} \rangle =\sum_{{\bm m} \in {\mathbb Z}^d} e^{ \mathrm{i} \pi ({\bm m}, \hat{{\bm B}}{\bm n})} f(\mathfrak{t}^{-1}_{\bm n}\omega, {\bm m}-{\bm n})|{\bm m}\rangle,
\end{equation}
where $|{\bm n}\rangle$ and $|{\bm m}\rangle$ should be viewed here as column matrices with entries $|{\bm m},\alpha\rangle$ and $|{\bm m},\beta\rangle$. The result of a representation $\{\pi_\omega f\}_{\omega \in \Omega}$ is automatically a covariant family of operators. For example, the Hamiltonian of Eq.~\ref{LatticeHamO} is generated by the element:
\begin{equation}
h_{\alpha, \beta}(\omega,{\bm n})=t_{\bm n}^{\alpha \beta} (\omega).
\end{equation}

If one introduces the norm on ${\cal A}_0$:
\begin{equation}\label{Norm}
\|f\| = \sup_{\omega \in \Omega} \|\pi_\omega f \|,
\end{equation}
and the $\ast$-operation: 
\begin{equation}
f^\ast(\omega,{\bm n})=f(\mathfrak{t}_{\bm n}^{-1}\omega,-{\bm n})^\dagger,
\end{equation}
then the completion of ${\cal A}_0$ under the norm of Eq.~\ref{Norm} becomes a $C^\ast$-algebra, which will be denoted by ${\cal A}$. The $C^\ast$-algebra structure allows one to develop the spectral problem and the functional calculus in a purely algebraic framework, without any reference to wavefunctions. The algebra ${\cal A}$ is the smallest algebra that one can consider and still be able to compute the response functions of the homogeneous systems.

\subsection{The non-commutative differential calculus}

The classical differential manifolds are completely determined by the algebra of smooth functions defined over the manifold and by the differential calculus with these functions. In the non-commutative setting, one replaces the commutative algebra of smooth functions over the classic Brillouin torus with the non-commutative algebra ${\cal A}$, and then builds a non-commutative differential calculus over ${\cal A}$. For the latter, one replaces ${\bm k}$-integration over the classical Brillouin torus by a trace over ${\cal A}$:
\begin{equation}\label{Trace}
{\cal T}(f)=\int_\Omega dP(\omega) \ \mathrm{tr} \{f(\omega,{\bm 0})\},
\end{equation}
where $\mathrm{tr}$ is the trace over ${\cal M}_{D \times D}$ space, and the ${\bm k}$-derivations by linear automorphisms of algebra ${\cal A}$:
\begin{equation}
(\partial_j f)(\omega,{\bm n}) = \mathrm{i} n_j f(\omega,{\bm n}), \ j=1,\ldots,d.
\end{equation}
Then the triplet $({\cal A},{\cal T},\partial)$ determines a non-commutative manifold called the non-commutative Brillouin torus.

The trace ${\cal T}$ has the standard properties:
\begin{enumerate}
\item Positivity: ${\cal T}(f*f^*) \ge 0$,
\item Cyclic-city: ${\cal T}(f*g)={\cal T}(g*f)$.
\end{enumerate}
For the translational invariant case, ${\cal T}$ reduces to the ordinary ${\bm k}$-integration, and for the general case with disorder and a magnetic field, the physical meaning of ${\cal T}$ is that of the trace per unit volume:
\begin{equation}
\lim\limits_{\Lambda\rightarrow \infty}\frac{1}{|\Lambda|}\mathrm{Tr}_{\Lambda}\{(\pi_\omega f) (\pi_\omega g) \ldots \}={\cal T}(f*g\ldots),
\end{equation}
where $\mathrm{Tr}_\Lambda$ means the trace over the quantum states inside the box $\Lambda$. 

The derivations are not defined over the entire algebra ${\cal A}$ but only over the sub-algebra of ``differentiable" functions: 
\begin{equation}
{\cal C}^1({\cal A})=\{f\in {\cal A}, \ \|\partial_j f\|<\infty, \ j=1,2,3\}.
\end{equation}
 More general, the sub-algebra of $N$-th differentiable functions is defined as: 
 \begin{equation}
 {\cal C}^N({\cal A})=\{f\in {\cal A}, \ \|\partial_1^{\alpha_1} \partial_2^{\alpha_2} \partial_3^{\alpha_3}f\|<\infty, \ \alpha_1+\alpha_2+\alpha_3=N\}.
 \end{equation}
We will only work with ${\cal C}^\infty({\cal A})$ elements, so from now all elements will be assumed to be part of this sub-algebra. 

For convenience, we collect below a list of well-known rules of calculus ($j,k=1,2,3$):
\begin{enumerate}
\item The derivations commute: 
\begin{equation}
\partial_j \partial_k f=\partial_k \partial_j f.
\end{equation}
\item The derivations are *-derivations:
\begin{equation}
\partial_j (f^*)=(\partial_j f)^*.
\end{equation}
\item The derivations satisfy the Leibniz rule: 
\begin{equation}
\partial_j(f*g)=(\partial_j f)*g + f*(\partial_j g).
\end{equation}
\item The operator representations of the derivations are:
\begin{equation}
\pi_\omega (\partial_j f)= -\mathrm{i}[x_j, \pi_\omega f],
\end{equation}
where ${\bm x}=(x_1,\ldots,x_s)$ is the position operator.
\item For $f$ invertible in ${\cal A}$, then:
\begin{equation}\label{DiffId1}
\partial_j f^{-1}=-f^{-1} * (\partial_j f) * f^{-1}.
\end{equation} 
\item If $\Phi$ and $\Phi'$ are any differentiable functions on the complex plane, then \cite{BELLISSARD:1994xj}:
\begin{equation} \label{DiffId2}
{\cal T}(\Phi(f)*\partial_j \Phi'(f) ) = 0.
\end{equation}
In particular:
\begin{equation} \label{DiffId2a}
{\cal T}(\partial_j f ) = 0.
\end{equation}
If combined with the Leibnitz rule, one gets the partial integration rule:
\begin{equation} \label{DiffId2p}
{\cal T}(\partial_j f *g) = -{\cal T}(f*\partial_j g ).
\end{equation}
\item If $\Phi$ is an analytic function in a neighborhood of the unit circle and $u$ is any unitary element ($u^*=u^{-1}$), then \cite{Prodan:2009od}:
\begin{equation}\label{DiffId3}
{\cal T}(\Phi(u) \partial_j u) = a_{-1} {\cal T}(u^{-1}\partial_j u),
\end{equation}
where $a_{-1}$ is the coefficient of $z^{-1}$ in the Laurent expansion $\Phi(z)=\sum_{n=-\infty}^\infty a_n z^n$.
\end{enumerate}

\section{The Ito derivative with respect to the magnetic field}\label{Ito}

As we already mentioned, the traditional linear response calculus with respect to the magnetic field can be very difficult. In the context of the magneto-electric response, this was already highlighted in Ref.~\cite{ChenPRB2011vb}. In the non-commutative setting, however, this task is quite straightforward, in great part due to the pioneering work by Rammal and Bellissard \cite{RammalJPF1990ki}. According to this work, the derivative with respect to ${\bm B}$ of the expected values can be computed as:
\begin{equation}
\partial_{B_j}{\cal T}(f) = {\cal T}(\delta_j f), \ \ j=1,2,3,
\end{equation}
where $\delta_j$'s are the Ito derivatives, which can be computed using the following rules of calculus (with $j$ taken mod 3):
\begin{equation}\label{C1}
\delta_j \partial_k f = \partial_k \delta_j f,
\end{equation}
\begin{equation}\label{C2}
\delta_j f^\ast =(\delta_j f)^\ast,
\end{equation}
\begin{eqnarray}\label{C3}
\delta_j(f*g)=(\delta_j f)*g + f*(\delta_j g) \\
   \indent \indent + \frac{\mathrm{i}}{2}(\partial_{j+1} f * \partial_{j+2} g - \partial_{j+2}f * \partial_{j+1}g),\nonumber
\end{eqnarray}
\begin{eqnarray}\label{C4}
\delta_j(f^{-1})=-f^{-1} *\delta_j f * f^{-1} \\
\indent \indent +\frac{\mathrm{i}}{2} f^{-1}*\left [ \partial_{j+1}f * f ^{-1},\partial_{j+2}f *f^{-1}\right ]. \nonumber
\end{eqnarray}
These rules of calculus are enough to compute any response function involving the magnetic field. For example \cite{Schulz-BaldesArxiv2012gh}:\medskip

\noindent{\bf Proposition 1.} If $p$ is an idempotent element ($p*p=p$), such as the projector onto the occupied states for our model, then:
\begin{equation}\label{I1}
p*(\delta_j p)*p=-\frac{\mathrm{i}}{2}p*[ \partial_{j+1}p,\partial_{j+2}p ]*p
\end{equation}
and
\begin{equation}\label{I2}
(1-p)*(\delta_j p)*(1-p)=\frac{\mathrm{i}}{2}(1-p)*[ \partial_{j+1}p,\partial_{j+2}p ]*(1-p).
\end{equation}
{\it Proof.} According to the rule in Eq.~\ref{C3}
\begin{equation}
\delta_j p = \delta_j (p*p)=(\delta_j p)*p + p*(\delta_j p) + \frac{\mathrm{i}}{2}[\partial_{j+1} p , \partial_{j+2} p ]
\end{equation}
Then Eq.~\ref{I1} follows after one multiplies the above equation with $p$ on both left and right sides and collects all three resulting $p*\delta_jp * p$ terms. For the second identity, note that $(1-p)$ is also an idempotent, so we can replace $p$ by $(1-p)$ in the first identity and use the fact that $\delta_j(1-p)=-\delta_jp$.\qed

\medskip The off-diagonal components of $\delta_j p$ do not have an expression that involves only $p$ and its derivatives like in Eqs.~\ref{I1} and \ref{I2}. If $p$ is the spectral projector of the Hamiltonian $h$ onto the spectrum below the insulating gap:
\begin{equation}
p=\frac{\mathrm{i}}{2\pi} \oint (h-z)^{-1}dz
\end{equation}
where the contour encircles the occupied spectrum of $h$, then one will compute the off-diagonal part of $\delta_j p$ using the rule in Eq.~\ref{C4} on $(h-z)^{-1}$. The result will depend on the particularities of $h$ and will involve cross-gap matrix elements. In fact, in general, if we are dealing with a (analytic) function of the Hamiltonian, $\Phi(h)$, such as the Fermi-Dirac function at finite temperatures, then we can use the representation:
\begin{equation}
\Phi(h)=\frac{\mathrm{i}}{2\pi} \oint \Phi(z) (h-z)^{-1}dz
\end{equation}
and the rule in Eq.~\ref{C4} on $(h-z)^{-1}$, to compute:
\begin{eqnarray}
\delta_j \Phi(h)  =\frac{\mathrm{i}}{2\pi} \oint dz \ \Phi(z) \left \{- (h-z)^{-1}*\delta_j h *(h-z)^{-1} \frac{ }{ }\right .\nonumber \\
\indent \left .+\frac{\mathrm{i}}{2}(h-z)^{-1}\left [\partial_{j+1}h*(h-z)^{-1},\partial_{j+2}h*(h-z)^{-1} \right ]\right \}.
\end{eqnarray}

\section{The isotropic magneto-electric response}\label{ME}

The magneto-electric tensor is well defined only when referenced from a ``standard" system. We will denote the reference system by $h_0$ and we will consider adiabatic time variations ({\it i.e.} infinitely slow) of the Hamiltonian $h(t)$ which originate at this $h_0$ and terminate at some $h$. The main assumption is that the Fermi level is always in a spectral gap during the adiabatic variations. At the technical level, one needs the assumption that $\partial_t h(t)=0$ at the beginning and at the end of the adiabatic deformation. Now, let $p(t)$ be the projector onto the states below this spectral gap:
\begin{equation}
p(t)=\big(1-\chi( h(t) - \epsilon_F)\big)/2,
\end{equation}
where $\chi(x)=\mathrm{sign}(x)$. Under these conditions, Ref.~\cite{Schulz-BaldesArxiv2012gh} established the following result.

\medskip \noindent {\bf Proposition 2.} The change of the electric polarization during an adiabatic variation  can be computed via the formula:
\begin{equation}\label{ElectricPolarization}
\Delta P_j=\mathrm{i}\int dt \ {\cal T}\big (p*[\partial_t p, \partial_j p] \big),
\end{equation}
where the $t$-dependence of $p$ is implied.\medskip

It is worthwhile to add a few more lines about the work of Ref.~\cite{Schulz-BaldesArxiv2012gh}. For the context of disordered systems under magnetic fields, these results are important because they lift the (super-) adiabatic theorem \cite{Nenciu:1993uq}, including the full non-adiabatic error estimation, to the non-commutative Brillouin torus. As such, one can foresee many additional applications which previously have been available only for the periodic case. Another notable point is that the formula for orbital magnetization derived in Ref.~\cite{Schulz-BaldesArxiv2012gh} does not require a spectral gap but only a mobility gap at the Fermi level. As such, there is hope that the theory of the magneto-electric response can be pushed into the regime of strong disorder. This is extremely important for the reasons to be discussed in the next sections.

Starting from the result stated in Proposition 1, we shall prove the following statement.\medskip

\noindent {\bf Proposition 3.} In the above conditions, the variation in the isotropic magneto-electric response:
\begin{equation}
\Delta \alpha = \frac{1}{3}\sum_{j=1}^{3} \partial \Delta P_j / \partial B_j
\end{equation}
 can be computed via :
\begin{eqnarray}\label{MainFormula}
\Delta \alpha =\frac{1}{2} \int dt  \ \varepsilon_{\alpha \beta \gamma \delta}{\cal T}\big ( p*\partial_\alpha p*\partial_\beta p*\partial_\gamma p *\partial_\delta p \big ) \nonumber \\
\indent  \indent +\frac{\mathrm{i}}{3} {\cal T}\big(\chi(h-\epsilon_F)*\partial_j p* \delta_j p \big ) \Big |_{\mathrm{initial}}^{\mathrm{final}},
\end{eqnarray}
where summation over the repeating indices is assumed. The indices $\alpha$, ..., $\delta$ run through $t$ and $j=1,2,3$, and $\varepsilon$ is the signature of the permutation.

\medskip \noindent {\bf Remark.} The first (second) term on the right hand side of Eq.~\ref{MainFormula} is called topological (non-topological) for reasons to be discussed shortly. When the magnetic field and disorder are turned off, the topological part of $\Delta \alpha$ reduces to the formula derived in Refs.~\cite{QiPRB2008ng,EssinPRB2010ls,MalashevichNJP2010bv} in the context of perfectly periodic insulators. Showing a similar correspondence for the non-topological term is a more involved task and is deferred to future investigations.

\medskip \noindent {\it Proof.} Since:
\begin{equation}\label{DAlpha}
\Delta \alpha = \mathrm{i}\int dt \ \frac{1}{3}\sum_{j=1}^3 {\cal T}\big (\delta_j(p*[\partial_t p, \partial_j p])\big ),
\end{equation}
our main task is to compute:
\begin{equation}
\sum_{j=1}^3 {\cal T}\big (\delta_j(p*[\partial_t p, \partial_j p])\big ).
\end{equation}
Using the rule of calculus from Eq.~\ref{C3}, we have:
\begin{eqnarray}\label{P1}
\delta_j(p*[\partial_t p, \partial_j p])=\delta_j p *[\partial p,\partial_j p]+p*\delta_j[\partial_t p,\partial_j p] \nonumber \\
\indent +\frac{\mathrm{i}}{2}\partial_{j+1}p*\partial_{j+2}[\partial_tp,\partial_jp]-\frac{\mathrm{i}}{2}\partial_{j+2}p*\partial_{j+1}[\partial_t p,\partial_j p].
\end{eqnarray}
The contributions to $\Delta\alpha$ from the last two terms above identically cancel out, as it can be easily seen by using an integration by parts (see Eq.~\ref{C2}). The Ito derivative of the commutator is:
\begin{eqnarray}\label{Partial1}
\delta_j[\partial_t p,\partial_j p ]&= [\delta_j \partial_t p,\partial_j p ] + [\partial_t p, \delta_j \partial_j p]    \nonumber \\
&  +\frac{\mathrm{i}}{2}\big(  [\partial_{j+1}\partial_t p,\partial_{j+2}\partial_j p ] - [\partial_{j+2}\partial_t p,\partial_{j+1}\partial_j p ]\big) .
 \label{Trace2}
\end{eqnarray} 
The contribution to the integrant in Eq.~\ref{DAlpha} from the first two terms of Eq.~\ref{Partial1} is:
\begin{eqnarray}
\indent \sum_{j=1}^3 {\cal T}\big (p*([\delta_j \partial_t p,\partial_j p ] + [\partial_t p, \delta_j \partial_j p]) \big ) \nonumber \\
= 2 \sum_{j=1}^3 {\cal T}\big ( [\partial_t p, \partial_j p]*\delta p_j \big) + \frac{d}{dt}\sum_{j=1}^3 {\cal T}\big (\chi(h-\epsilon_F)*\partial_j p*\delta_j p\big ).
\end{eqnarray}
Indeed:
\begin{eqnarray}
 \indent \mathcal{T} \big(p * [\delta_j \partial_t p,\partial_j p ]+p*[\partial_t p, \delta_j \partial_j p] \big)  \\ 
 =\mathcal{T} \big(p*\delta_j\partial_t p*\partial_j p -p*\partial_j p*\delta_j\partial_t p +p*\partial_t p*\delta_j \partial_j p - p*\delta_j\partial_jp*\partial_t p \big) \nonumber \\
 =  \mathcal{T} \big(\partial_j p *p*\delta_j\partial_t p-p*\partial_j p*\delta_j \partial_t p +p*\partial_t p*\delta_j \partial_j p - \partial_t p *p * \delta_j\partial_j p\big)
\nonumber
\\
=  \mathcal{T} \big(\partial_t(\partial_j p *p*\delta_j p)-\partial_t\partial_j p *p *\delta_jp-\partial_j p*\partial_tp*\delta_j p \nonumber \\
\indent -\partial_t(p*\partial_jp*\delta_j p)+\partial_t p*\partial_j p*\delta_jp + p*\partial_t\partial_jp*\delta_j p\nonumber \\
\indent -\partial_j(p*\partial_tp)*\delta_j p +\partial_j(\partial_t p * p)* \delta_j p\big) \nonumber \\
=\partial_t \mathcal{T}\big((\partial_j p *p-p*\partial_jp)*\delta_j p\big )+2{\cal T}\big((\partial_tp*\partial_jp-\partial_j p *\partial_tp)*\delta_jp\big), \nonumber
\end{eqnarray} 
and the affirmation follows because $\partial_jp*p=(1-p)*\partial_jp$.

The contribution to the integrant in Eq.~\ref{DAlpha} from the last two terms of Eq.~\ref{Partial1} is zero:
\begin{equation}
\sum_{j=1}^3 {\cal T}\left (p*\big(  [\partial_{j+1}\partial_t p,\partial_{j+2}\partial_j p ] - [\partial_{j+2}\partial_t p,\partial_{j+1}\partial_j p ]\big)\right ) =0.
\end{equation}
Indeed, after an integration by parts, this contribution can be written as:
\begin{eqnarray}
  \sum_{j=1}^3 \mathcal{T}  \big(-\partial_{j+1}p*[\partial_t p,\partial_{j+2}\partial_j p ] 
 -p * [\partial_{t},\partial_{j+1} \partial_{j+2}\partial_j p ] \nonumber \\
 \indent +\partial_{j+2} p * [\partial_t p, \partial_{j+1} \partial_j p] +p * [\partial_t,\partial_{j+2}\partial_{j+1}\partial_j p] \big)  \nonumber  \\
 =\sum_{j=1}^3 \mathcal{T}   \big (-\partial_{j+1}p * [\partial_t p,\partial_{j+2}\partial_j p ] +\partial_{j+2} p * [\partial_t p, \partial_{j+1} \partial_j p] \big ) \nonumber \\
 = \mathcal{T} \big(-\partial_2 p * [\partial_tp,\partial_3 \partial_1 p] +\partial_3 p * [\partial_t p,\partial_2 \partial_1 p]  \nonumber \\
\indent -\partial_3 p * [\partial_tp,\partial_1 \partial_2 p] +\partial_1 p * [\partial_tp,\partial_3 \partial_2 p]  \nonumber  \\
 \indent  -\partial_1 p * [\partial_t,\partial_2 \partial_3 p] +\partial_2 p * [\partial_t,\partial_1 \partial_3 p] \big ),
\end{eqnarray}
and in the final lines one can easily identify pairs of terms that identically cancel out.

 So far the calculation stands as:
\begin{eqnarray}
\sum_{j=1}^3 {\cal T}\big (\delta_j(p*[\partial_t p, \partial_j p])\big)=
3\sum_{j=1}^3{\cal T}\big([\partial_t p ,\partial_j p]*\delta_jp\big) \nonumber \\
\indent+\frac{d}{dt}\sum_{j=1}^3 {\cal T}\big (\chi(h-\epsilon_F)*\partial_j p*\delta_j p\big )
\end{eqnarray}
Now:
\begin{eqnarray}
\indent \mathcal{T} \big([\partial_tp, \partial_j p]*\delta_j p\big) \\
 =\mathcal{T} \big( [\partial_tp, \partial_j p]*\delta_j p*p) + \mathcal{T} \big( [\partial_tp, \partial_j p] *\delta_jp*(1-p)\big) \nonumber \\
 =\mathcal{T} \big( p *[\partial_t p, \partial_j p]*\delta_j p*p) + \mathcal{T} \big( (1-p) *[\partial_t p, \partial_j p]*\delta_j p*(1-p)\big) \nonumber, 
\end{eqnarray}
where in the last line we used the cyclic-city of the trace and the idempotency of $p$ and $1-p$. Given the fact that $p$ and $(1-p)$ commute with $[ \partial_tp, \partial_jp]$, we can continue as:
\begin{eqnarray}
\ldots= \mathcal{T}\big([ \partial_t p, \partial_j p]*(p*\delta_jp*p+(1-p)*\delta_jp*(1-p))\big)
\end{eqnarray}
at which point we can use the rules in Eqs.~\ref{I1} and \ref{I2} to continue as:
\begin{eqnarray}
\ldots= \mathcal{T}\big([ \partial_t p, \partial_j p]*(-\frac{\mathrm{i}}{2}p*[ \partial_{j+1}p,\partial_{j+2}p ]*p \nonumber \\
\indent \indent +\frac{\mathrm{i}}{2}(1-p)*[ \partial_{j+1}p,\partial_{j+2}p ]*(1-p))\big) \nonumber \\
=\frac{1}{2\mathrm{i}}\mathcal{T}\big((2p-1)*[ \partial_t p, \partial_j p]*[ \partial_{j+1}p,\partial_{j+2}p ]\big).
\end{eqnarray}
It remains to demonstrate that:
\begin{eqnarray}
\sum_{j=1}^3\mathcal{T}\big([ \partial_t p, \partial_j p]*[ \partial_{j+1}p,\partial_{j+2}p ]\big)=0
\end{eqnarray}
which follows after an explicit display of the terms involved in the sum:
\begin{eqnarray}
\indent \sum_{j=1}^3\mathcal{T}\big([ \partial_t p, \partial_j p]*[ \partial_{j+1}p,\partial_{j+2}p ]\big)\nonumber \\
=\mathcal{T}\big( \partial_t p *\partial_1 p*\partial_2 p*\partial_3 p -\partial_t p *\partial_1 p*\partial_3 p*\partial_2 p \nonumber \\
\indent \indent - \partial_1 p *\partial_t p*\partial_2 p*\partial_3 p +\partial_1 p *\partial_t p*\partial_3 p*\partial_2 p \nonumber \\ 
\indent +\partial_t p *\partial_2 p*\partial_3 p*\partial_1 p -\partial_t p *\partial_2 p*\partial_1 p*\partial_3 p \nonumber \\
\indent \indent - \partial_2 p *\partial_t p*\partial_3 p*\partial_1 p +\partial_2 p *\partial_t p*\partial_1 p*\partial_3 p \nonumber \\
\indent +\partial_t p *\partial_3 p*\partial_1 p*\partial_2 p -\partial_t p *\partial_3 p*\partial_2 p*\partial_1 p \nonumber \\
\indent \indent - \partial_3 p *\partial_t p*\partial_1 p*\partial_2 p +\partial_3 p *\partial_t p*\partial_2 p*\partial_1 p \big )
\end{eqnarray}
and using the cyclic-city of the trace one can easily identify the pairs of terms that identically cancel out. At this point we arrived at:
\begin{eqnarray}
\sum_{j=1}^3 {\cal T}\big (\delta_j(p*[\partial_t p, \partial_j p])\big)=
\frac{3}{\mathrm{i}}\sum_{j=1}^3 \mathcal{T}\big(p*[ \partial_t p, \partial_j p]*[ \partial_{j+1}p,\partial_{j+2}p ]\big) \nonumber \\
\indent+\frac{d}{dt}\sum_{j=1}^3 {\cal T}\big (\chi(h-\epsilon_F)*\partial_j p*\delta_j p\big ),
\end{eqnarray}
and the first term on the left side can be anti-symmetrized and brought to the desired form. This concludes the proof.\qed

\section{Quantization and Invariance}

\begin{figure}
\center
  \includegraphics[width=8cm]{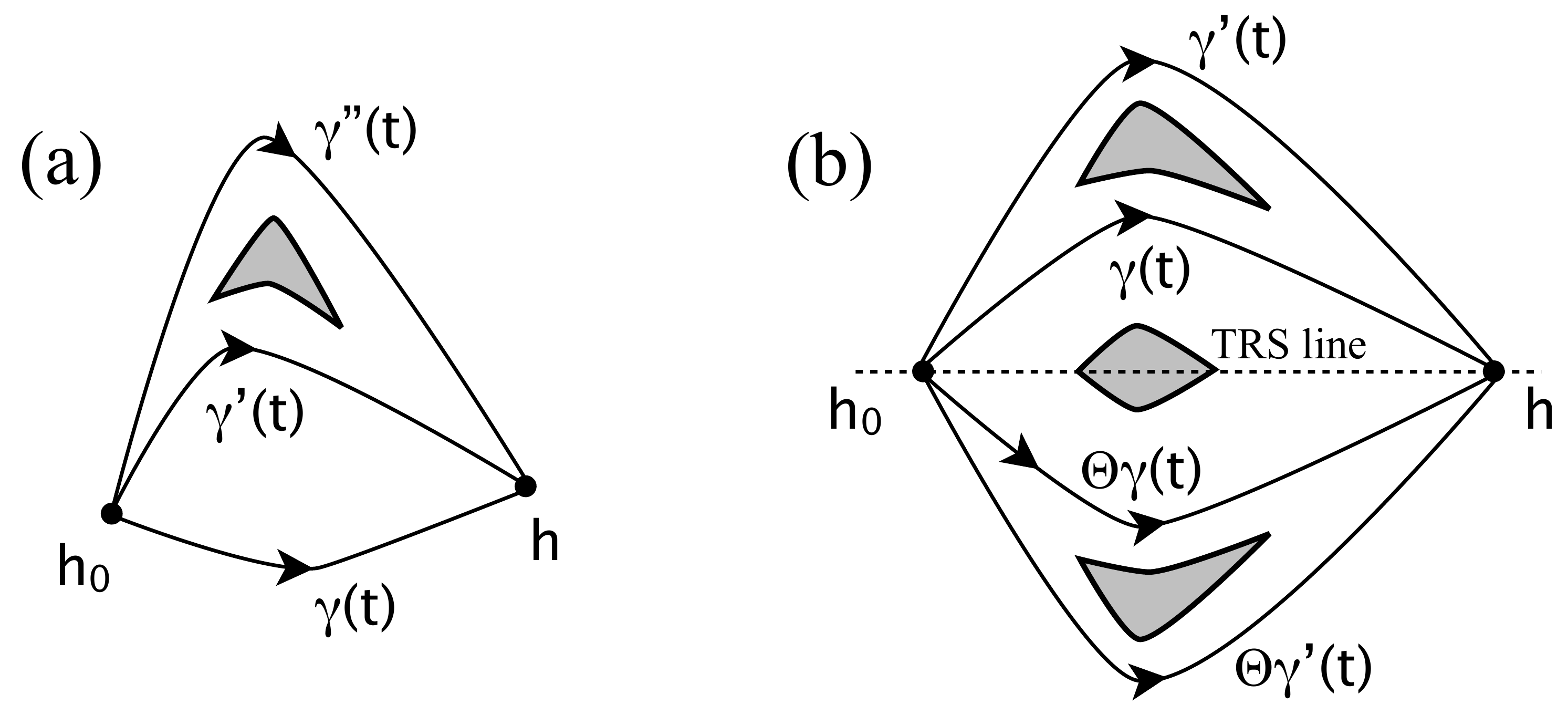}\\
  \caption{(a) Various adiabatic interpolations between the initial ($h_0$) and final ($h$) systems. (b) Here $h_0$ and $h$ are time-reversal symmetric (TRS), but they are adiabatically interpolated via paths that are not necessarily time-reversal symmetric ($\Theta \gamma \neq \gamma$). The shaded regions represents regions in the model's parameter space where the spectral gap closes. The TRS line is invariant to the time-reversal operation.}
 \label{ApproxDiagram}
\end{figure} 

For the purpose of this discussion, it is more convenient to view the adiabatic time evolution from an initial to a final configuration as an adiabatic interpolation between two points in the parameter space of the model. The adiabatic interpolations will be represented as curves in this parameter space and the time-integral can be viewed as an integral along such curve. We automatically assume that the spectral gap at the Fermi level remains open during the adiabatic interpolations. We will use the notations $\gamma+\gamma'$ to denote the interpolation path obtained by joining the interpolation paths $\gamma$ and $\gamma'$, $-\gamma$ to denote the interpolation path $\gamma$ when taken in reverse, and $\Delta \alpha[\gamma]$ to indicate that the variation in the magneto-electric response depends on the interpolation path $\gamma$.

The first question we want to address is what happens with the variation of the isotropic magneto-electric response when one considers different interpolation paths, $\gamma(t)$ and $\gamma'(t)$, between the reference and the final systems. Since:
\begin{equation}
\Delta \alpha[\gamma]-\Delta \alpha[\gamma']=\Delta \alpha[\gamma-\gamma'],
\end{equation}
we can write:
\begin{eqnarray}
\Delta \alpha[\gamma] -\Delta \alpha[\gamma'] 
 =\frac{1}{2}\oint\limits_ {\gamma - \gamma'}dt  \ \varepsilon_{\alpha \beta \gamma \delta}{\cal T}\big( p*\partial_\alpha p*\partial_\beta p*\partial_\gamma p *\partial_\delta p \big ),
\end{eqnarray}
and the important observation here is that the $t$-integral is now taken over a closed loop. One can then recognize on the right-hand side the expression of the second Chern number $C_2$ over the manifold $(\gamma - \gamma') \times \mbox{Brillouin torus}$, written in the real-space rather than in the $k$-space. Because of the finite-gap assumption, the second Chern number is quantized, a fact that follows from a general argument by Connes \cite{CONNES:1985cc}, which says that the pairing between the projectors and cyclic cocycles is a homotopy invariant. Hence the quantization occurring in the clean limit persists when the disorder is turned on, at least as long as the spectral gap remains open.

Therefore, the answer to our first question is that the difference in $\Delta \alpha[\gamma]$, when computed along different adiabatic evolutions, is always an integer number (we recall that our units were taken such that $e=h=1$). We illustrate in Fig.~1(a) various possible situations. For example, if the interpolating paths $\gamma(t)$ and $\gamma'(t)$ can be deformed into each other without closing the spectral gap, then $C_2$ is necessarily zero and $\Delta \alpha[\gamma]$ is identical for the two interpolation paths. But, if the interpolating paths cannot be deformed into each other without closing the spectral gap, such as it happens for $\gamma(t)$ and $\gamma''(t)$ in Fig.~1(a), then $C_2$ may take non-trivial integer values so $\Delta \alpha[\gamma]$ and $\Delta \alpha[\gamma'']$ cannot be expected to be identical, but we can say with absolute confidence that the difference between the two will be an integer number.

We now turn our attention to adiabatic interpolations between time-reversal invariant systems, in which case we need to turn the magnetic field off. Symmetry considerations can be used to show that only the topological part of $\Delta \alpha$ survives for such such systems \cite{EssinPRB2010ls,MalashevichNJP2010bv}. The time-reversal operation is given by the map:
\begin{equation}
\Theta: {\cal A}|_{{\bm B}=0}\rightarrow {\cal A}|_{{\bm B}=0}, \ \ \ (\Theta f)(\omega,{\bm n}) = e^{\mathrm{i}  \pi s_y} \overline{f(\omega,{\bm n})}e^{-\mathrm{i} \pi s_y},
\end{equation}
where $s_y$ is a matrix from ${\cal M}_{D\times D}$ representing the $y$-component of the spin operator, and the line at the top represents complex conjugation. Some immediate properties of time-reversal map $\Theta$ are:
\begin{equation}\label{V1}
\Theta(f*g)=(\Theta f)*(\Theta g),
\end{equation}
and
\begin{equation}\label{V2}
\partial_j \Theta f = - \Theta (\partial_j f).
\end{equation} 

Now, let $h_0$ and $h$ be two time-reversal invariant Hamiltonians, $\Theta h_0= h_0$ and $\Theta h = h$. The key observation is that, if $\gamma(t)$ is an adiabatic interpolation between $h_0$ and $h$, then $\Theta \gamma$ is also an adiabatic interpolation between $h_0$ and $h$ (with a strictly positive spectral gap), and furthermore:
\begin{equation}\label{V3}
\Delta \alpha[\gamma] = \Delta \alpha [-\Theta \gamma].
\end{equation}
This can be seen from the explicit formula of Eq.~\ref{MainFormula} and the properties of the time-reversal map stated in Eqs.~\ref{V1} and \ref{V2}. We already established that:
\begin{eqnarray}
\Delta \alpha[\gamma]-\Delta \alpha[\Theta \gamma] =C_2[\gamma-\Theta \gamma],
\end{eqnarray}
and this together with Eq.~\ref{V3} leads to an important conclusion for this section:
\begin{equation}
\Delta \alpha [\gamma]=\frac{1}{2}C_2[\gamma-\Theta \gamma].
\end{equation}
This is a generalization of the Qi-Hughes-Zhang argument \cite{QiPRB2008ng} to the case of disordered insulators.

In Fig.~1(b) we illustrate several possible situations. If $\gamma$ can be deformed to a time-reversal invariant path, $\Theta \gamma = \gamma$, then $C_2=0$ and $\Delta \alpha[\gamma]$ is necessary zero. But if $h_0$ and $h$ cannot be connected with a time-reversal invariant path without closing the spectral gap at the Fermi level, then $\gamma$ must be taken as in Fig.~1(b) so that to avoid the closing of the spectral gap. In this case,  $C_2[\gamma - \Theta \gamma]$ can take non-trivial integer values and consequently $\Delta \alpha[\gamma]$ can take non-trivial values, which are either integer or half-integer numbers. If we take a different interpolation path, such as $\gamma'$ in Fig.~1(b), then $\Delta \alpha[\gamma']$ may change but only by an integer number, as it was already established above. As such, for all possible adiabatic interpolation paths, $\Delta \alpha[\gamma]$ is either an integer number or a half-integer number. In other words, the integer or half-integer character of $\Delta \alpha[\gamma]$ is topologically stable. 

\section{Conclusions} 

The results of the last section can be further developed into a $\mathbb{Z}_2$ classification of the 3-dimensional disordered time-reversal invariant insulators. For this, one can take the reference systems to be the insulators with zero hopping amplitudes between different lattice sites. These systems can be called ``trivial" since they do not display any magneto-electric response as there are no Peierls factors when a magnetic field is turned on. As such, we can postulate $\alpha=0$ for the trivial systems. Then, the time-reversal invariant insulators fall into two classes (here we assume that all time-reversal symmetric insulators can be adiabatically connected to a representative from the trivial class): one for which $\alpha = \mathrm{integer}$ and one for which $\alpha = \mbox{half-integer}$, and we can say with absolute confidence that an insulator from the first class and an insulator from the second class cannot be connected by a time-reversal invariant interpolation path without closing the insulating gap. An insulator from a given symmetry class is generally called topological \cite{Hughes2010gh} if it cannot be adiabatically connected with its atomic limit (obtained by turning off the hopping terms between distinct lattice sites) via an interpolation path that respects the symmetry of the class and without closing the insulating gap. If we follow this definition, then the class of 3-dimensional disordered time-reversal invariant insulators with $\alpha= \mbox{half-integer}$ contains only topological insulators. Of course, our analysis does not rule out additional, more refined topological sub-classifications of either $\alpha= \mbox{half}$ or $\alpha= \mbox{half-integer}$ classes.

A $\mathbb{Z}_2$ classification of the disordered inversion symmetric insulators is also possible via similar arguments. This is the case because the inversion operation satisfies the properties written in Eqs.~\ref{V1} and \ref{V2}, which are at the core of the $\mathbb{Z}_2$ classification of time-reversal insulators. However, for inversion symmetric insulators, it is explicitly known \cite{Hughes2010gh} that the $\mathbb{Z}_2$ classification is not the end of the story as it can be furtherly refined.

We would like to point out that the computation of Section~\ref{ME} remains valid even in the regime of strong disorder where the insulating gap is closed and is replaced by a mobility gap (provided we accept the formula Eq.~\ref{ElectricPolarization} for the electric polarization). If a non-commutative theory of the second Chern number paralleling that of the first Chern number \cite{BELLISSARD:1994xj} can be developed, then the magneto-electric response can be used to topologically classify the time-reversal and inversion symmetric insulators in the strong disorder regime, which is one of the long-standing open problems in the field.

The non-commutative formula Eq.~\ref{MainFormula} of the magneto-electric response can be efficiently evaluated on a computer by using the methods developed in Ref.~\cite{ProdanAMRX2012bn}. As such, we now have a practical tool to explore, at least numerically, the physics of strongly disordered 3-dimensional topological insulators. The previous numerical attempts \cite{LeungPRB2012vb} to compute the 3-dimensional $\mathbb{Z}_2$ invariants based on twisted boundary methods were limited to small systems due to the extreme complexity of such methods, even though these calculations involved a gauge invariant, highly efficient formulation of the $\mathbb{Z}_2$ invariant \cite{ProdanPRB2011vy}. From our experience with other non-commutative formulas (such as those for the first Chern number and for the Kubo formula \cite{ProdanJPhysA2011xk,ProdanAMRX2012bn,Xue2012fh}), we are quite optimistic that the non-commutative formula for magneto-electric polarization will enabled computations for much larger volumes where finer and more accurate results can be obtained.

\ack We thank David Vanderbilt for extremely useful discussions. We also thank the reviewers for extremely useful comments. This work was supported by the U.S. NSF grants DMS-1066045 and DMR-1056168. 

\bigskip

\noindent {\bf References}\medskip

\bibliographystyle{iopart-num}
 

\providecommand{\newblock}{}

\end{document}